\documentclass[runningheads]{llncs}

\newif\ifsubmit
\newif\ifextended
\newif\iflualatex

\submittrue
\extendedtrue
\lualatexfalse

\usepackage[T1]{fontenc}
\usepackage{graphicx}
\usepackage[pdfpagelabels=true]{hyperref}

\usepackage{xcolor}

\usepackage{amsmath}
\usepackage{amssymb}
\usepackage{breqn}
\usepackage{cite}

\usepackage{xspace}
\usepackage{newunicodechar}
\iflualatex
	\usepackage{emoji}
\else
	\usepackage{hwemoji}
\fi
\usepackage{pifont} 
\usepackage[frozencache,cachedir=minted-cache]{minted}
\usepackage{paralist}
\usepackage{tabularx}
\usepackage{nicefrac}
\usepackage{booktabs}
\usepackage{caption}
\usepackage{subcaption}
\usepackage{cleveref}
\usepackage{tikz}

\usetikzlibrary{backgrounds,patterns,patterns.meta,arrows,arrows.meta,calc,shapes,shadows,decorations.pathmorphing,decorations.pathreplacing,automata,shapes.multipart,positioning,shapes.geometric,fit,circuits,trees,shapes.gates.logic.US,fit, shadows.blur,shapes.symbols,intersections}
\usepackage{pgfplots}
\pgfplotsset{
    compat=1.18,
    legend style={
        nodes={anchor=base},
        name=leg
    },
    legend image post style={yshift=.5ex},
    every legend to name picture/.style={
        baseline={(leg.base)},
    },
}

\crefname{equation}{Eq.}{Eq.}
\crefname{pluralequation}{Eqs.}{Eqs.}

\crefname{figure}{Fig.}{Fig.}
\crefname{pluralfigure}{Figs.}{Figs.}

\crefname{section}{Sect.}{Sect.}
\crefname{pluralsection}{Sects.}{Sects.}

\crefname{appendix}{App.}{App.}
\crefname{pluralappendix}{Apps.}{Apps.}

\crefname{table}{Tab.}{Tab.}
\crefname{pluraltable}{Tabs.}{Tabs.}

\crefname{definition}{Def.}{Def.}
\crefname{pluraldefinition}{Defs.}{Defs.}

\crefname{theorem}{Theorem}{Theorems}
\crefname{pluraltheorem}{Theorems}{Theorems}

\crefname{lemma}{Lemma}{Lemma}
\crefname{plurallemma}{Lemmas}{Lemmas}

\crefname{example}{Example}{Example}
\crefname{pluralexample}{Examples}{Examples}

\crefname{assumption}{Assumption}{Assumption}
\crefname{pluralassumption}{Assumptions}{Assumptions}

\crefname{remark}{Remark}{Remark}
\crefname{pluralremark}{Remarks}{Remarks}

\crefname{listing}{Listing}{Listing}
\crefname{plurallisting}{Listings}{Listings}

\definecolor{red}{rgb}{0.745,0.192,0.102}
\definecolor{darkgreen}{RGB}{34,161,55}

\newcommand{\ie}{i.e.\@\xspace}
\newcommand{\wrt}{w.r.t.\@\xspace}
\newcommand{\eg}{e.g.,\@\xspace}
\newcommand{\cf}{cf.\@\xspace}

\newcommand{\M}{\ensuremath{\mathcal{M}}}
\newcommand{\Act}{\ensuremath{\textit{Act}}}
\newcommand{\Pf}{\ensuremath{\mathbf{P}}}
\newcommand{\AP}{\ensuremath{\textit{AP}}}
\newcommand{\Prob}{\ensuremath{\mathbb{P}}}
\newcommand{\Rew}{\mathbb{R}}
\newcommand{\pol}{\ensuremath{\sigma}}
\newcommand{\Distr}[1]{\mathsf{Distr}(#1)}

\newcommand{\tool}[1]{\textsc{#1}\xspace}
\newcommand{\storm}{\tool{Storm}}
\newcommand{\stormpy}{\tool{stormpy}}
\newcommand{\stormvogel}{\tool{Stormvogel}}
\newcommand{\bird}{\tool{bird}}
\newcommand{\dtcontrol}{\tool{dtControl}}
\newcommand{\dtmap}{\tool{dtMap}}
\newcommand{\paynt}{\tool{PAYNT}}
\newcommand{\prism}{\tool{Prism}}
\newcommand{\jani}{\tool{JANI}}
\newcommand{\modest}{\tool{Modest}}
\newcommand{\mrmc}{\tool{MRMC}}

\ifthenelse{\boolean{lualatex}}{%
	\newunicodechar{🍏}{\gapple{}}
	\newunicodechar{🍐}{\pear{}}
	\newunicodechar{🍒}{\cherries{}}
	\newunicodechar{🍇}{\grapes{}}
	\newunicodechar{🐦}{\raven{}}
	\newunicodechar{⬛}{}
	\newunicodechar{🍉}{\fruit{}}
	\newunicodechar{🧺}{\basket{}}
	\newunicodechar{🎲}{\emoji{game-die}}

	\newcommand{\gapple}{\emoji{green-apple}}
	\newcommand{\pear}{\emoji{pear}}
	\newcommand{\cherries}{\emoji{cherries}}
	\newcommand{\grapes}{\emoji{grapes}}
	\newcommand{\raven}{\emoji{bird}}
	\newcommand{\basket}{\emoji{basket}}
	\newcommand{\fruit}{\emoji{watermelon}}
	\newcommand{\die}{\emoji{game-die}}
}{
	\newcommand{\gapple}{🍏}
	\newcommand{\pear}{🍐}
	\newcommand{\cherries}{🍒}
	\newcommand{\grapes}{🍇}
	\newcommand{\raven}{🐦}
	\newcommand{\basket}{🧺}
	\newcommand{\fruit}{🍉}
	\newcommand{\die}{🎲}
}

\newcommand{\codefilecustom}[4]{\inputminted[linenos,firstline=#3,lastline=#4,fontsize=\footnotesize]{#2}{#1}}
\newcommand{\codefile}[3]{\codefilecustom{#1}{python}{#2}{#3}}
\newmintinline[code]{python}{escapeinside=||}
\newminted[pycon]{pycon}{fontsize=\footnotesize,escapeinside=||}
\newcommand{\lstvspace}{\vspace{-12pt}}

\newcommand\smallpar[1]{%
\medskip\noindent\emph{#1}}

\newcommand\smallsubsubsection[1]{%
\medskip\noindent\textbf{#1}}

\makeatletter
\def\orcidID#1{\smash{\href{http://orcid.org/#1}{\protect\raisebox{-1.25pt}{\protect\includegraphics{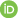}}}}}
\makeatother

\begin{document}

\title{Probabilistic Model Checking Taken by Storm}
\subtitle{A Tutorial on the Probabilistic Model Checker Storm}
\author{Matthias Volk\inst{1}\orcidID{0000-0002-3810-4185}
\and Linus Heck\inst{2}\orcidID{0000-0002-4774-7609}
\and Sebastian Junges\inst{2}\orcidID{0000-0003-0978-8466}
\and\\ Joost-Pieter Katoen\inst{3}\orcidID{0000-0002-6143-1926}
\and Tim Quatmann\inst{3}\orcidID{0000-0002-2843-5511}}
\authorrunning{M. Volk et al.}
\institute{
Eindhoven University of Technology, Eindhoven, The Netherlands
\and Radboud University, Nijmegen, The Netherlands
\and RWTH Aachen University, Aachen, Germany}

\maketitle

\begin{abstract}
This tutorial paper presents a hands-on perspective on probabilistic model checking with the Storm model checker. 
Storm is a decade-old model checker that excels in performance and a rich Python-based ecosystem, which makes it easy to integrate in various workflows.
This tutorial focuses on Markov decision processes (MDP), which are popular in a variety of fields.
It demonstrates the basic workflow, from Python-based modeling, model checking with a variety of properties, to the extraction of policies.
Further, it showcases the support for recent topics that focus on different types of uncertainty, such as interval MDP and POMDP,
and the ability to quickly implement simple algorithms on top of existing data structures.
\end{abstract}

\section{What is Probabilistic Model Checking?}
\label{sec:intro}
The verification of safety-critical systems requires faithfully modeling uncertainty about system behavior, whether it is due to noisy sensors, hardware degradation, or randomization as used in distributed protocols. The use of probabilistic models has developed as a promising direction to support verification of the aforementioned systems~\cite{DBLP:journals/arcras/KwiatkowskaN022}. 
In particular, \emph{probabilistic model checking~(PMC)}, an extension of classical (non-probabilistic) model checking~\cite{BK08}, allows to verify whether a model with probabilistic behavior satisfies the given requirements.
Given a Markov model and a formal property of interest, (classical) PMC checks whether the probability of the model satisfying the property is above (or below) a given threshold and returns the yes/no answer.
Moreover, probabilistic model checkers can also compute the probability---up to some precision guarantee---and therefore answer queries such as \textit{``what is the maximal probability of collecting all resources before a (randomly determined) deadline''} or \textit{``what is the expected time to collect all resources?''}

\begin{figure}[t]
\centering
\input{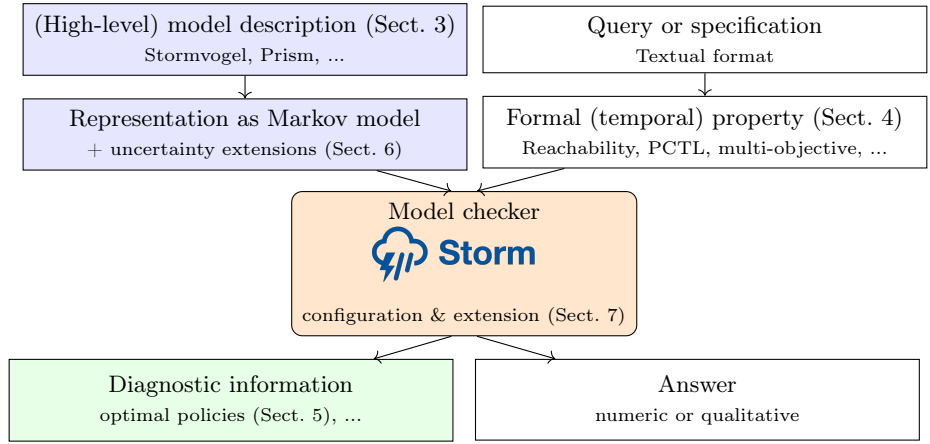}%
\caption{Probabilistic model checking (extended from~\cite{DBLP:conf/birthday/HenselJQ025} with tutorial-specific content in small font)}
\label{fig:overview}
\end{figure}

Historically, there has been a variety of probabilistic model checkers, including~E-(MC)$^2$~\cite{DBLP:conf/tacas/HermannsKMS00}, \mrmc~\cite{DBLP:journals/pe/KatoenZHHJ11}, \tool{Modest-mcsta}~\cite{DBLP:conf/tacas/HartmannsH14}, \prism\cite{DBLP:conf/cav/KwiatkowskaNP11}, \tool{EPMC} / \tool{iscasMC}~\cite{DBLP:conf/fm/HahnLSTZ14}, \storm~\cite{DBLP:journals/sttt/HenselJKQV22}, \tool{UPPAAL}~\cite{DBLP:journals/sttt/LarsenPY97}, and \text{PET}~\cite{DBLP:conf/atva/Meggendorfer22}.
Application areas of PMC are surveyed in~\cite{DBLP:conf/birthday/KwiatkowskaN025} and the capabilities of PMC tools have been presented 
for various tools, such as \prism{}~\cite{DBLP:conf/sfm/ForejtKNP11}, \modest~\cite{DBLP:journals/corr/abs-2203-09881} and \storm{}~\cite{DBLP:conf/birthday/HenselJQ025}.

The classical workflow of probabilistic model checking is depicted in \cref{fig:overview} and consists of the following three main steps in sequence:
\begin{compactenum}
    \item \emph{Model building} (indicated in blue color in \cref{fig:overview}).
    First, the model is specified precisely. The de-facto standard are high-level model descriptions, using, \eg the \modest language~\cite{DBLP:journals/tse/BohnenkampDHK06}, the \prism language~\cite{DBLP:conf/cav/KwiatkowskaNP11}, or the \jani interchange format~\cite{DBLP:conf/tacas/BuddeDHHJT17}.
    From this high-level description of the model, the underlying explicit Markov model---including transition matrices and state labels---is automatically created.
    Alternatively, the underlying model can already be  specified explicitly, typically originating from external model construction.
    \item \emph{Model checking} (orange color).
    Using preprocessing steps, the model is reduced and put into a standard form.
    In that form, the model checker must solve a set of, \eg Bellman equations\cite{DBLP:books/wi/Puterman94}, using highly optimized routines. 
    Solving these equations typically translates into an answer of the original query on the original model.
    \item \emph{Diagnostic information} (green color).
    In addition to the qualitative or numerical answer, further diagnostic information can be produced, although it may induce significant overhead. 
\end{compactenum}

Modern usage of probabilistic model checking typically goes beyond such linear workflows. The different steps become more and more intertwined and embedded into reasoning loops in an effort to solve real problems. For example, in addition to the answer---\textit{``what is the winning probability''?}---diagnostic information yields the optimal strategy and for this optimal strategy, additional queries can then be used to refine, \eg the original model. 

\smallsubsubsection{Storm.}
The development of such loops in the PMC workflow is eased by the ability to do rapid prototyping in Python. The open-source model checker \storm{}~\cite{DBLP:journals/sttt/HenselJKQV22} has long excelled in performance~\cite{DBLP:conf/isola/BuddeHKKPQTZ20}.
Recent advances and tools in its ecosystem now provide a rich API to help researchers in various branches of formal methods to use probabilistic model checking in their own tools and methods.
We refer to \storm's website \href{https://www.stormchecker.org/}{www.stormchecker.org} for general information.

\smallsubsubsection{This tutorial.}
We will present the various facets of PMC for \emph{Markov decision processes (MDPs)} combining both probabilistic and nondeterministic behavior.
The tutorial demonstrates \storm's functionality by means of the running example of the \emph{Orchard} board game, a conceptually simple children's game. 

The tutorial is aimed towards users who want to make first steps applying PMC.
We assume general familiarity with MDPs and recap relevant notation in \cref{sec:prelim}.
We refer to~\cite{DBLP:conf/lics/Katoen16} for an introduction into PMC and to~\cite{BK08,DBLP:reference/mc/BaierAFK18} for the relevant background on Markov models and model checking algorithms.

\smallsubsubsection{Outline.}
The tutorial will cover the aspects depicted in \cref{fig:overview}.
In the first step, \cref{sec:modeling}, we demonstrate how to model the board game in a high-level manner.
In the second step, \cref{sec:model_checking}, we specify and check a variety of properties, such as reachability probabilities, rewards and multi-objective queries.
In the third step, \cref{sec:policies}, we use the diagnostic information provided by \storm to obtain \eg the optimal strategies for winning the Orchard game.
In the fourth step, \cref{sec:uncertainty}, we show how our (classical) MDP model can be extended by including additional sources of uncertainty.
We cover \emph{interval MDP} which allow uncertainty in the transition probabilities, and \emph{partially observable MDP (POMDP)} which allow uncertainty about the exact state by means of restricted observations.
Lastly, in \cref{sec:next_steps}, we demonstrate how the \storm model checker can be configured to user needs, and how to extend \storm by writing simple Python algorithms.

\smallsubsubsection{Data.}
We provide the models, code examples and instructions of this tutorial as supplementary material online\footnote{\url{https://doi.org/10.5281/zenodo.18955665}}.
These Jupyter notebooks can be executed interactively alongside this tutorial and allow to easily explore the \storm ecosystem.
\ifthenelse{\boolean{extended}}{}{%
    The full code examples are also available in the extended version~\cite{extended}.
}

\section{Preliminaries}
\label{sec:prelim}
We briefly recap Markov decision processes, policies, and properties. An introduction to the topic can be found in~\cite{BK08,DBLP:reference/mc/BaierAFK18,DBLP:conf/tacas/HartmannsJQW23}. In the tutorial presentation, we will elaborate these definitions.

\smallsubsubsection{MDPs.}
We model our game as a \emph{Markov decision process (MDP)} which combines probabilistic and nondeterministic behavior. Let $\Distr{X}$ denote distributions over finite sets $X$.

\begin{definition}[Markov decision process (MDP)]
	\label{def:mdp}
	An \emph{MDP} is a tuple $\M = (S, s_0, \Act, \Pf, \AP, L)$ with
	a finite set $S$ of \emph{states} and an initial state $s_0$, 
	a finite set   $\Act$ of \emph{actions}, a partial transition function  $\Pf\colon S \times \Act \nrightarrow  \Distr{S}$ mapping state-action pairs to distributions over successors,
	and a finite set $\AP$ of \emph{atomic propositions}  with a state labeling function $L\colon S \to 2^{\AP}$.
\end{definition}
The partial transition function clarifies that not all actions are necessarily enabled in all states.
In particular, the set of \emph{enabled actions} in a state $s$ is given by
$\Act(s) = \{\alpha \in \Act \mid \Pf(s,\alpha) \text{ is defined} \}$. A \emph{choice} is a pair of state and enabled action.
As is customary, we write $\Pf(s,\alpha,s')$ for $\Pf(s,\alpha)(s')$ if defined and assume $\Pf(s,\alpha,s')=0$ otherwise. Thus, an MDP transitions from state $s \in S$ with action $\alpha \in Act$ to state $s' \in S$ with probability $\Pf(s,\alpha,s')$.
A \emph{discrete-time Markov chain (DTMC)} is an MDP with one choice per state.

\smallsubsubsection{Rewards.}
An MDP can be naturally extended with \emph{rewards} (also called costs).
We use action-rewards here which associate a non-negative reward $R(s,\alpha) \in \mathbb{R}_{\geq0}$ with each pair of state $s$ and enabled action $\alpha \in \Act(s)$.
Visiting an action accumulates the associated rewards. 
An MDP can have multiple reward structures.

\smallsubsubsection{Policies.}
A \emph{policy} (also called \emph{strategy} or \emph{scheduler}) on an MDP $\M$ is a function $\pol\colon (S \times \Act)^{*} \times S \to \Act$ mapping paths to actions, such that $\pol(\dots  \cdot s) \in \Act(s)$. A special case of such (deterministic) policies are \emph{memoryless} policies where the action choice depends solely on the last state. Policies resolve nondeterminism. 
A (memoryless) policy $\pol$ on an MDP $\M$ \emph{induces} a DTMC $\M^\pol$ by applying the policy in each state.
The probability measure $\text{Pr}^{\M^\pol}_s$ on an MDP $\M$ is defined via the given policy $\pol$. Similar constructions can be made for general policies.

\smallsubsubsection{Specifications.}
Probabilistic model checkers use (various extensions to) \emph{probabilistic computation tree logic (PCTL)} or \emph{linear temporal logic (LTL)} to express properties of interest~\cite{BK08,DBLP:reference/mc/BaierAFK18}. The standard queries quantify universally, or existentially, over policies and ask, \eg whether the probability to satisfy a temporal path property is above a threshold for every policy. The probability to satisfy a path property for a specific policy is defined by a standard probability measure on the DTMC induced by the MDP and the policy. 

\section{Step One: Modeling}
\label{sec:modeling}

We use a simple board game as a running example throughout this tutorial.
Board games typically involve probabilistic behaviour, \eg by throwing dice, and nondeterministic behavior based on the players' choices.
The analysis of a board game typically involves optimizing the players' strategies.
Board games are therefore a good model for demonstrating the various modeling and model checking capabilities of \storm.

\subsection{The Orchard Game}
\label{subsec:orchard}
The Orchard game (also called Obstgarten, Le verger, Boomgaard, El Frutal, Første Frukthage) is a simple children's game.
In the following, we model the \textit{First Orchard} game variant based on the official rule book available online.\footnote{\url{https://cdn.hff.de/image/upload/v1/22/30/28/14/22302814.pdf}}

The game is played cooperatively.
There are four types of \emph{fruit}---apples~\gapple{}, pears~\pear{}, cherries~\cherries{} and plums~\grapes{}\footnote{We use the grape symbol as there is no plum symbol.}.
For each type of fruit, there is a tree with four pieces of this fruit.
There is a \emph{basket} \basket{} for collecting fruit.
As antagonist, there is a \emph{raven}~\raven{} which reaches the orchard after five steps.
The goal of the game is to collect all fruit before the raven reaches the orchard. In each round, a player throws a six-sided dice and performs an action depending on its outcome:
\begin{compactitem}
	\item If the outcome is a type of fruit~\gapple{}\pear{}\cherries{}\grapes{}, then the player picks a piece of this type from the  tree and places it in the basket. 
	If no fruit is left on the tree, the player cannot pick anything.
	\item If the outcome is the fruit basket~\basket{}, then the player may pick any fruit.
	\item If the outcome is the raven~\raven{}, the raven moves one step towards the orchard.
\end{compactitem}
The players win iff they collect all fruit before the raven arrives at the orchard.

\subsection{The Orchard Model}
Before starting the actual modeling, we make some general modeling decisions.
First, we decide which parts are relevant for the model.
In the Orchard game, we need to keep track of the remaining fruit on the trees and the position of the raven.
As the game is cooperative, it is irrelevant how many players are playing and we assume one player.
We also do not need to keep track of the basket as only the remaining fruit on the trees are relevant for the winning condition.

In our Orchard model, we divide each round into two distinct phases: first throwing the dice and then acting on the outcome.
This allows to separate the probabilistic dice throw from the player's (potentially nondeterministic) action.

Lastly, to facilitate analysis of different variants of the game, we parameterize the model in terms of fruit types, number of fruit and distance of the raven.

\subsection{Defining the Orchard MDP in Stormvogel}
\label{subsec:model_declaration}
We use the \stormvogel library~\cite{Stormvogel25} for our initial model declaration.
\stormvogel\footnote{\url{https://github.com/stormchecker/stormvogel}} is a Python library providing easy-to-use interfaces for probabilistic model checking.
A main feature of \stormvogel is the \bird API, which allows to specify the model using Python code.
This enables quickly declaring models by using the familiar Python language instead of a dedicated modeling language with a steeper learning curve. 
While the API allows natural modeling, the user must still pay attention in order to obtain small and finite models.

\smallsubsubsection{Defining an MDP in Stormvogel.}
To create an MDP model in \stormvogel, we need to provide all ingredients of the tuple $\M = (S, s_0, \Act, \Pf, \AP, L)$.
Starting from the initial state $s_0$ of the MDP, the \bird API of \stormvogel automatically generates the successor states by following the transition function $\Pf$ for each enabled action $\Act(s)$.
Exploring all newly created states allows to subsequently generate the complete state space $S$ of the MDP and \stormvogel keeps track of all states found so far.
If a successor state already exists, it does not need to be explored further.
The state space exploration terminates once all states have been explored.
The check whether a state already exists is based on the \emph{equality function} of a state.
Two Python state objects which are equal \wrt the equality function must represent the same (logical) state and the equality function must consider all relevant information to distinguish different states.

It is the responsibility of the user to ensure that the state space exploration terminates, \ie that the state space is finite.
In our Orchard example, the game will eventually end and therefore only finitely many distinct states are explored.
In general, there might be infinitely many states that would need exploration---for instance due to unbounded variables like counters, lists, etc.---potentially resulting in non-termination of the \bird API.
\stormvogel prevents such issue by using a configurable threshold \code{max_size} on the explored state space.

\smallsubsubsection{State representation.}
We start the modelling by defining general data structures for the Orchard game, the different types of fruit, the possible outcomes of the dice and the game end.
\ifthenelse{\boolean{extended}}{%
	The corresponding Python code is given in \cref{lst:stormvogel_data_structures} in the appendix.
}{%
	The corresponding Python code can be found in~\cite{extended}.
}%

We introduce a class \code{Orchard} which represents the current state of the game, see \cref{lst:stormvogel_game_class_extract}.
\ifthenelse{\boolean{extended}}{%
	The full class is given in \cref{lst:stormvogel_game_class} in the appendix.
}{%
	The full class can be found in in~\cite{extended}.
}%
\begin{listing}[t]
	\codefile{code/orchard_game_stormvogel.py}{38}{43}
	\lstvspace
	\caption{An extract of the Orchard game class.}
	\label{lst:stormvogel_game_class_extract}
\end{listing}
This class inherits from the \stormvogel \code{State} class.
The \code{Orchard} object is initialized with a list of configuration parameters such as the considered types of fruit, the number of fruit per tree, and the distance of the raven.
The game initializes the variable \code{trees} which keeps track of the remaining number of fruit per tree.
It also keeps track of the outcome of the \code{dice}, \ie \basket{}, \raven{} or a fruit.
In case of a fruit, the second entry of the \code{dice} tuple denotes the specific type of fruit which was thrown.
Value \code{None} represents that the dice needs to be thrown first.

The \code{Orchard} class also defines various functions for ease of modeling.
For instance, \code{game_state} returns whether the game is ongoing or who has won.
Function \code{pick_fruit} picks the given fruit from the tree (if available).
Function \code{next_round} resets the dice for the next round, \code{move_raven} moves the raven.
Modeling these functions separately also yields a modular design which allows to easily modify specific behavior, such as the movement of the raven.

\smallsubsubsection{Initial state.}
The initial state $s_0$ is given by the initialization of our \code{Orchard} class.
At first, we use a smaller configuration of the game with only two different types of fruit, two pieces of fruit per tree, and a distance of two for the raven.
\begin{pycon}
>>> init_small = Orchard([Fruit.APPLE, Fruit.CHERRY],
...                       num_fruits=2, raven_distance=2)
>>> print(init_small)
2|\gapple|, 2|\cherries{}|, 2|$\leftarrow$\raven|
\end{pycon}
Printing the initial state of the simplified game shows that there are two apples and two cherries on the trees, and the raven is two steps away from its goal.

\smallsubsubsection{Available actions.}
The transition function returning successor states requires information which actions are enabled in each state.
\begin{listing}[t]
	\codefile{code/orchard_game_stormvogel.py}{89}{107}
	\lstvspace
	\caption{The available actions.}
	\label{lst:stormvogel_available_actions_extract}
\end{listing}
\Cref{lst:stormvogel_available_actions_extract} defines $\Act(s)$ for any state $s$.
In the Orchard game, we use the following actions:
\begin{compactitem}
	\item \code{gameEnded} {\footnotesize(ll.~91--92)}: the only action available once the game has ended.
		This action represents the self-loop for such final states.
	\item \code{nextRound} {\footnotesize(ll.~93--94)}: the dice must be thrown again.
	\item \code{pickF} {(\footnotesize ll.~95--96)}: the dice outcome is $\mathtt{F} {\in} \{$\gapple{}$,$\pear{}$,$\cherries{}$,$\grapes{}$\}$ and the player picks the corresponding fruit \code{F} from the tree.
	\item \code{chooseF} {\footnotesize(ll.~97--105)}: dice outcome \basket{} and fruit $\mathtt{F}{\in}\{$\gapple{}$,$\pear{}$,$\cherries{}$,$\grapes{}$\}$ is chosen.
	\item \code{moveRaven} {\footnotesize(ll.~106--107)}: dice outcome \raven{} and the raven must be moved next.
\end{compactitem}
In most states, a single action is available.
The only exception is dice outcome~\basket{}, for which we list all fruit which are still available for choosing by the player.

\smallsubsubsection{Transition function.}
\stormvogel defines the transition function~$\Pf$ via function \code{delta} which takes as arguments a state and an available action, and returns the distribution over successor states as a sparse list of nonzero transition probabilities for different target states.
We define the transition function for the Orchard game in \cref{lst:stormvogel_transitions_extract}.
\begin{listing}[tp]
	\codefile{code/orchard_game_stormvogel.py}{111}{162}
	\lstvspace
	\caption{The transition function \code{delta}.}
	\label{lst:stormvogel_transitions_extract}
\end{listing}
If the game has ended, we add a self-loop to the same state with probability one {\footnotesize(ll.~113--115)}.
Otherwise, we perform each round in two phases: first throwing the dice and then acting on the outcome.
These are two separate transitions in the model.

In the first phase {\footnotesize(ll.~117--139)}, the dice outcome is \code{None} and the player must throw the dice.
We consider the outcomes for the different fruit types plus two additional outcomes \basket{} and \raven{}.
Each outcome has the same uniform probability \code{fair_dice_prob}, which is $\nicefrac{1}{|\textit{fruit_types}| + 2}$.
The rest of the first phase {\footnotesize(ll.~124--139)} creates the different successor states depending on the dice outcomes.
To this end, we create the \code{next_state} as a copy of the current state, and set the dice outcome \code{next_state.dice}.
Afterwards, we store the successor state together with the corresponding probability in the list \code{outcomes} of successor states.

The remainder of the function handles the second phase {\footnotesize(ll.~141--162)}, where the player acts depending on the dice outcome \code{state.dice}.
If the dice outcome is a fruit, the player performs \code{pick_fruit(fruit)} on the \code{next_state} and then finishes the turn with \code{next_round} {\footnotesize(ll.~141--147)}.
If the dice outcome is \basket{}, the player needs to choose a fruit {\footnotesize(ll.~149--156)}.
This choice is modeled by the different available actions \code{chooseF} in this state.
Based on the chosen \code{action} given as argument to the \code{delta} function, we can extract the chosen fruit.
For example, action name \code{chooseAPPLE} corresponds to \gapple{}.
In the successor state, we pick the chosen fruit and then continue with the next round.
Lastly, if the dice outcome is \raven{}, the raven moves {\footnotesize(ll.~158--162)}.

\smallsubsubsection{Labeling.}
We use a labeling $L$ to associate atomic propositions $\AP$ to states which satisfy them.
These labels help identify states which fulfill a certain property which is relevant to our analysis.
\ifthenelse{\boolean{extended}}{%
	The labeling for the Orchard game is given in \cref{lst:stormvogel_labels} in the appendix.
}{%
	The labeling for the Orchard game can be found in~\cite{extended}.
}%
The function \code{labels()} returns a list of atomic propositions per state.
As we are mainly interested in the winner of a game, we introduce two labels \code{PlayersWon} and \code{RavenWon} which mark each state with the party who won.
We also label each state with the string representation of the game state to help understanding and investigating the resulting model as the internals of a state become visible.
For scalable analysis, such kind of internal information is typically omitted in the final version of a model.

\smallsubsubsection{Rewards.}
Lastly, we specify a reward which counts the number of rounds.
\ifthenelse{\boolean{extended}}{%
	The reward function is given in \cref{lst:stormvogel_rewards} in the appendix.
}{%
	The reward function can be found in~\cite{extended}.
}%
Each state in which the dice is thrown has an action reward of one.
All other actions have reward zero.

\smallsubsubsection{Build model.}
All ingredients for the model have been specified and we can build the MDP model for the Orchard game.
\begin{pycon}
>>> orchard_simple = stormvogel.bird.build_bird(
...     modeltype=stormvogel.ModelType.MDP, init=init_small,
...     available_actions=available_actions, delta=delta,
...     labels=labels, rewards=rewards)
>>> print(orchard_simple.summary())
ModelType.MDP model with 90 states, 7 actions, and 33 distinct labels.
\end{pycon}
The function specifies the type of model---\code{MDP} here--and the MDP tuple entries.

We can also easily build the full model by configuring the initial state accordingly and calling \code{build_bird()} as before.
We need to increase the \code{max_size} of the explored state space as the resulting model has around $22,500$ states\footnote{\stormvogel has a small default \code{max_size} to quickly spot issues with toy examples used in teaching.}.
\begin{pycon}
>>> init_game = Orchard([Fruit.APPLE, Fruit.CHERRY, Fruit.PEAR, Fruit.PLUM],
...                      num_fruits=4, raven_distance=5)
>>> orchard = stormvogel.bird.build_bird(..., max_size=100000)
>>> print(orchard.summary())
ModelType.MDP model with 22469 states, 11 actions, and 3 distinct labels.
\end{pycon}
Some of the analyses in the following require a representation of the Orchard model in \storm data structures.
We can simply translate the model:
\begin{pycon}
>>> import stormpy
>>> orchard_storm = stormvogel.mapping.stormvogel_to_stormpy(orchard)
\end{pycon}

\subsection{Defining the Orchard MDP in the Prism Language}
\label{subsec:other_inputs}

We showed how to create an MDP model via \stormvogel.
\storm also supports creating models from other input formats such as the \prism language or the \jani interchange format.
In this tutorial, we also provide a \prism model for our Orchard game.
\ifthenelse{\boolean{extended}}{%
	The full specification is given in \Cref{app:orchard_prism}.
}{%
	The full specification can be found in~\cite{extended}.
}%
The \prism specification follows the same structure as the \stormvogel model.
It can be configured via global constants such as \code{NUM_FRUIT} and built via \stormpy.
The current game state is represented by variables such as \code{raven} or \code{apple}.
\begin{pycon}
>>> prism_program = stormpy.parse_prism_program('orchard_stormvogel.pm')
>>> constants = "NUM_FRUIT=4, DISTANCE_RAVEN=5"
>>> prism_program = stormpy.preprocess_symbolic_input(prism_program,
...                     [], constants)[0].as_prism_program()
>>> options = stormpy.BuilderOptions()
>>> options.set_build_state_valuations()
>>> options.set_build_choice_labels()
>>> options.set_build_all_labels()
>>> options.set_build_with_choice_origins()
>>> orchard_prism = stormpy.build_sparse_model_with_options(prism_program,
...                                                         options)
>>> print("Model with {} states and {} transitions".format(
...         orchard_prism.nr_states, orchard_prism.nr_transitions))
Model with 22469 states and 44954 transitions
\end{pycon}

In the remainder, \code{orchard_simple} refers to the  game with only two fruit types, \code{orchard} is the \stormvogel model of the full game, \code{orchard_storm} its corresponding representation in \storm, and \code{orchard_prism} is that model, built from the \prism specification.
The latter three are semantically equivalent.

\subsection{Model Inspection}
\label{subsec:model_inspection}
Before applying model checking, we inspect the model to spot modeling issues. We suggest to use a combination of simulations and visualizations. \stormvogel also allows to generate simulation runs through the model using random or user-provided policies and it can expose models via a gym-compatible interface.

\begin{figure}[t]
	\centering
	\includegraphics[height=6.5cm]{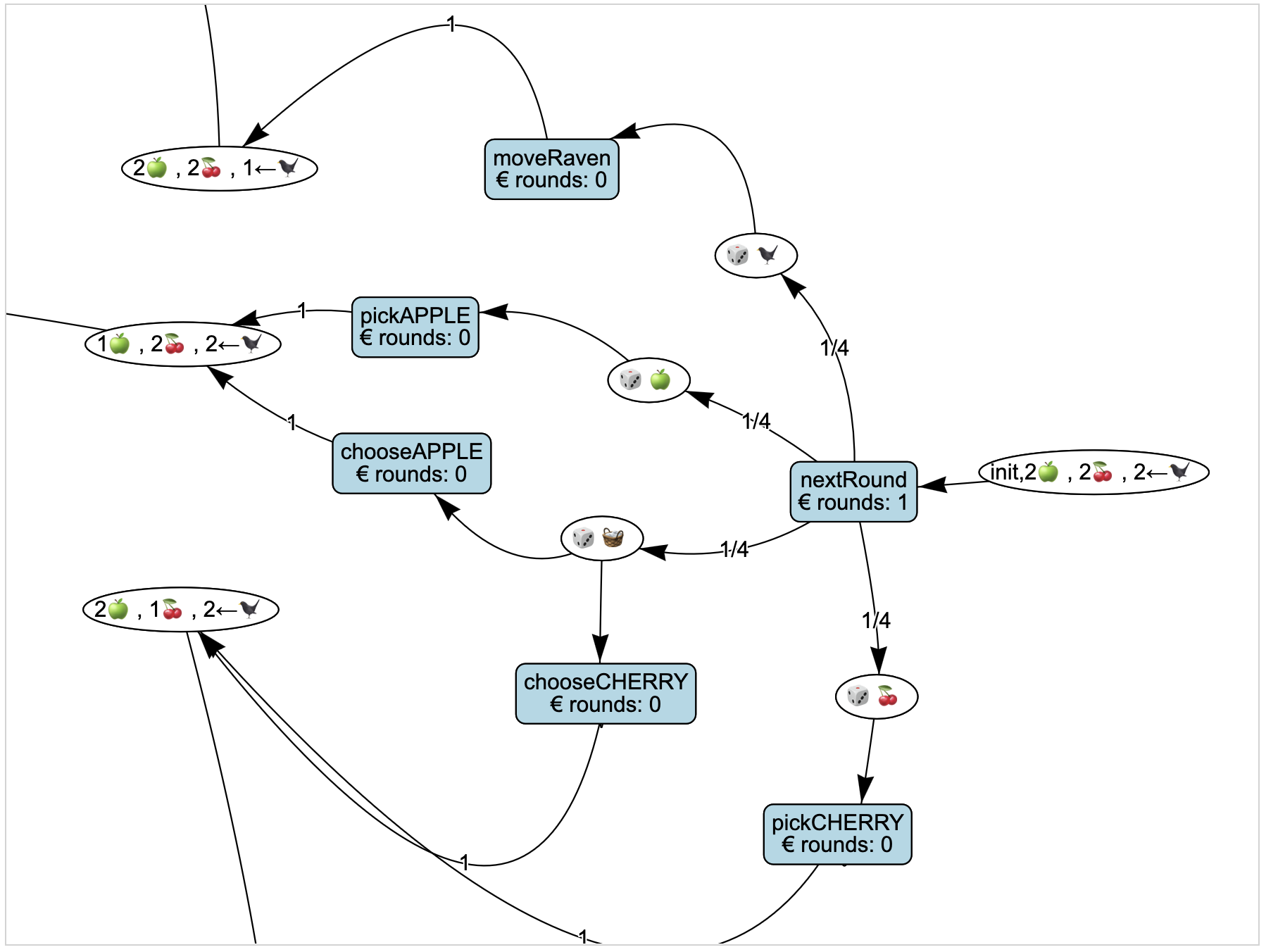}
	\caption{State space of the simplified MDP.}
	\label{fig:stormvogel_state_space}
\end{figure}

\stormvogel visualizes a model with \code{stormvogel.show(orchard_simple)}.
The interactive visualization is based on JavaScript and is therefore available via a web browser.
An extract of the resulting state space is shown in \cref{fig:stormvogel_state_space}.
States are depicted as ellipses, actions by blue boxes and transitions by arrows.
Labels are given inside the states and actions, and the rewards are indicated by the € symbol.
Starting from the initial state on the right, the only possible action is \code{nextRound}, which corresponds to throwing the dice.
It has reward one.
The four outcomes are depicted by the four successor states with symbol \die{} and are each reached with transition probability $\nicefrac{1}{4}$.
Afterwards, the corresponding action can be performed.
For example, from state \die{}\gapple{}, action \code{pickAPPLE} is possible and leads to a state 1\gapple{}, 2\cherries{}, 2$\leftarrow$\raven{} where only one apple is remaining. 
For larger state space sizes, \stormvogel has the option to  interactively \emph{explore} the state space by clicking on the successor states or by connecting it to the simulator.

\section{Step Two: Model Checking}
\label{sec:model_checking}
After creating and inspecting the model, we will now apply model checking to the Orchard MDP.
While, strictly speaking, model checking asks whether a property holds on an MDP, we will see that modern probabilistic model checking tools, including \storm, can actually go beyond such queries.
The section is structured based on the type of properties that we consider.

\subsection{Reachability}
\label{subsec:mc_reachability}
One of the simplest properties for MDPs is a reachability query \textit{``what is the maximal probability to reach a specific set of states, described by $\varphi$?''} and is denoted in probabilistic model checking as $\Prob_{\max=?}(F\,\varphi)$.
The probability operator $\Prob_{max=?}(\psi)$ denotes the maximal probability, ranging over all policies, that in the induced Markov chain, a path models a path formula $\psi$.
In this case $\psi := F\,\varphi$ (also $\lozenge \varphi$) represents finally reaching a state which satisfies a Boolean formula  $\varphi$ over atomic propositions~$\AP$. 
In \storm and in \stormvogel, we specify formulas in the de-facto standard \prism representation, \eg \code{'Pmax=? [F "Label"]'}.

In the Orchard game, we are mostly interested in the \textit{``maximal probability of winning the game''}.
\ifthenelse{\boolean{extended}}{%
	Using the label \code{PlayersWon} introduced in \cref{lst:stormvogel_labels},
}{%
	Using the label \code{PlayersWon},
}%
we express this property as $\Prob_{\max=?}(F\,\mathtt{PlayersWon})$, \ie maximizing the probability of reaching a state where the players have won. Notably, for many queries, we indeed get a result for every state and we therefore clarify in the code that we are interested in the probability from the initial state.
\begin{pycon}
>>> prob_players_won = 'Pmax=? [F "PlayersWon"]'
>>> result = stormvogel.model_checking(orchard_simple, prob_players_won)
>>> print(result.get_result_of_state(orchard_simple.get_initial_state()))
0.5711805425946498
\end{pycon}

\begin{figure}[t]
	\centering
	\includegraphics[height=7cm]{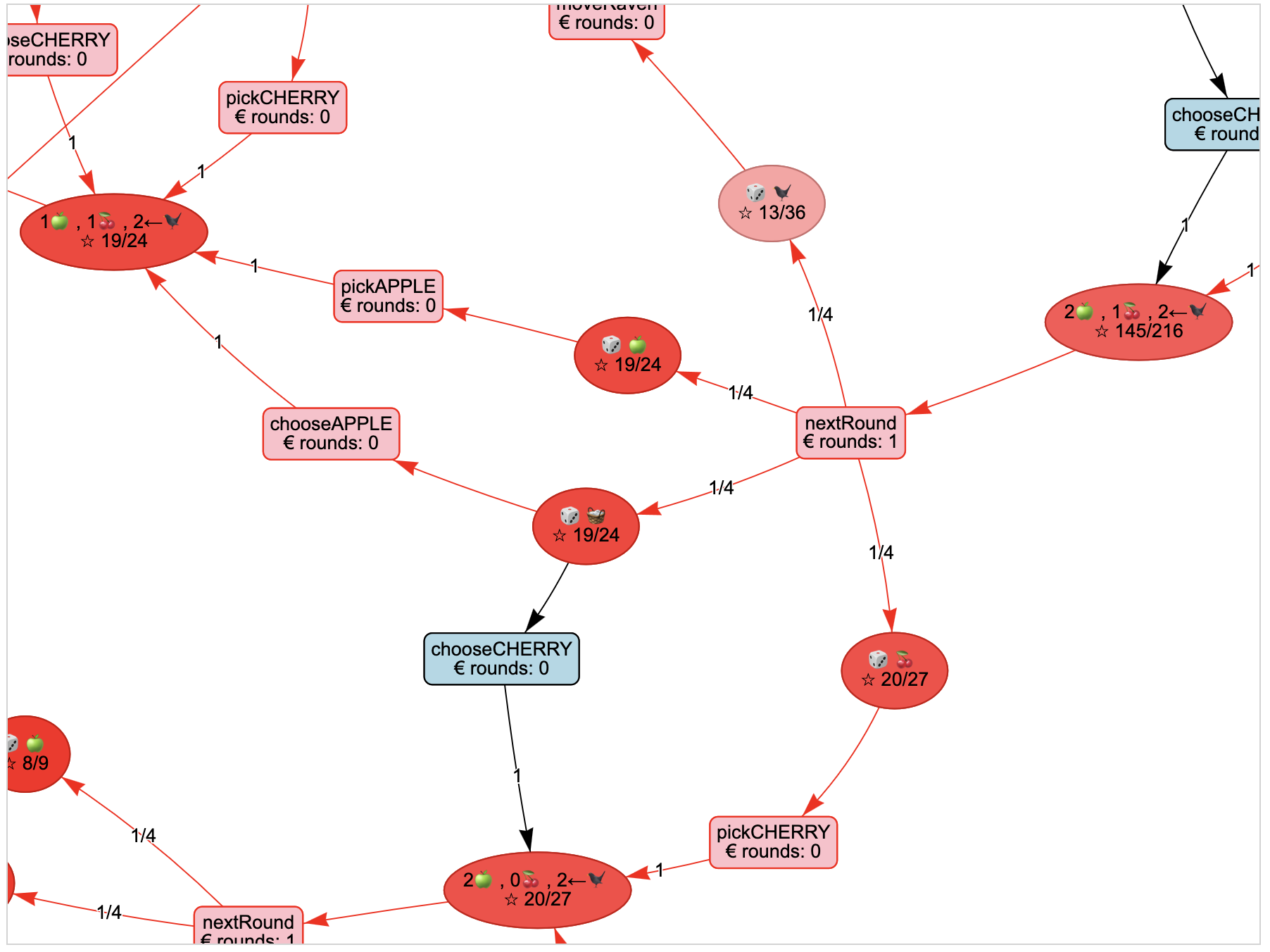}
	\caption{Visualization of the winning probability on the simplified MDP.}
	\label{fig:stormvogel_results}
\end{figure}

\noindent We can also visualize the result of this query on the MDP state space.
\begin{pycon}
>>> vis = stormvogel.show(orchard_simple, result=result)
\end{pycon}
An extract is depicted in \cref{fig:stormvogel_results}.
In each state, \ding{80} indicates the computed value from that state onward. The state is also colored with a red gradient depending on the result---the darker a state the higher its winning probability.
For instance, from the state 2\gapple{}, 1\cherries{}, 2$\leftarrow$\raven{} on the right, the players have a winning probability of $\nicefrac{145}{216}=0.671$, whereas from the \die{}\raven{}-successor, the probability is only $\nicefrac{13}{36}=0.3611$, as the raven is then one step closer to the orchard.

The model checking also returns a memoryless policy $\pol$ which ensures the maximal winning probability.
For each state $s$, the policy chooses one action $\pol(s)$.
This action is highlighted in red color in the \stormvogel visualization.
For example, from state \die{}\basket{}, the player has the option to either \code{chooseAPPLE} or \code{chooseCHERRY}.
The optimal policy chooses \gapple{}, because the successor state has a winning probability of $\nicefrac{19}{24}=0.7916$ which is higher than for \cherries{}, $\nicefrac{20}{27}=0.7407$.

For the full model \code{orchard}, the winning probability is $0.6314$.

\subsection{Total Rewards}
\label{subsec:mc_rewards}
Probabilistic model checkers commonly handle reward-based queries like \textit{``what is the maximal expected total reward until reaching a specific set of states?''} which can be expressed as $\Rew^{\mathit{rew}}_{\max=?}(F\,\varphi)$, where $\varphi$ is again describing a set of states and $\mathit{rew}$ refers to the name of a reward structure.

In the Orchard game, the \textit{``maximal expected total number of rounds until the game ends''} is described by $\Rew^{\mathit{rounds}}_{\max=?}\big(F\,(\mathtt{PlayersWon} \vee \mathtt{RavenWon})\big)$. Abbreviating the code to get the result for the initial state as above, we simply write: 
\begin{pycon}
>>> reward_prop = 'R{"rounds"}max=? [F "PlayersWon" | "RavenWon"]'
>>> print(stormvogel.model_checking(orchard, reward_prop).get_res[..])
22.339089182046724
>>> reward_prop = 'R{"rounds"}min=? [F "PlayersWon" | "RavenWon"]'
>>> print(stormvogel.model_checking(orchard, reward_prop).get_res[..])
20.882789624542582
\end{pycon}
Model checking reveals that the game lasts between 20.88 and 22.34 rounds on average---depending on the chosen strategy of the players.

\subsection{Beyond}
\label{subsec:mc_beyond}
\storm supports several queries that go significantly beyond the classical queries. While we used the \stormvogel model checking wrapper around \storm, for such properties we must often operate directly on the (somewhat more intricate) \stormpy API.
We use \Cref{lst:mc_helper} for simplicity.
\begin{listing}[t]
	\codefile{code/orchard_game_storm.py}{30}{33}
	\lstvspace
	\caption{Function for model checking calls.}
	\label{lst:mc_helper}
\end{listing}

\begin{figure}[t]
	\centering
	\includegraphics[height=4cm]{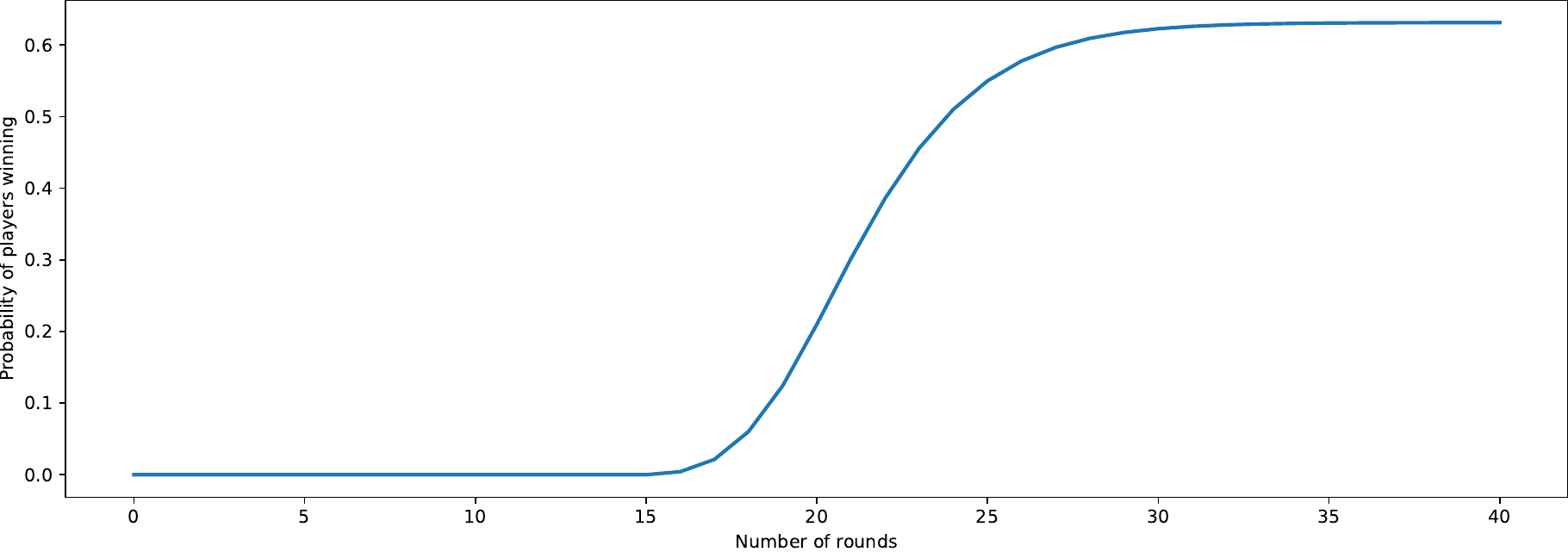}
	\caption{Winning probability within given number of rounds.}
	\label{fig:plot_win_rounds}
\end{figure}

\smallsubsubsection{Reward-bounded reachability probabilities~\cite{DBLP:journals/jar/HartmannsJKQ20}}
$\Prob_{\max=?}(F^{\mathit{rew} \le \lambda}\,\varphi)$ ask for the probability to reach a $\varphi$-state while having accumulated at most $\lambda$ reward.
In the Orchard game, we can compute the winning probability for varying number of rounds and plot the result in \cref{fig:plot_win_rounds}.
\begin{pycon}
>>> probabilities = []
>>> for k in range(41):
...     win_steps = 'Pmax=? [F{"rounds"}<=' + str(k) + ' "PlayersWon"]'
...     probabilities.append(model_check(orchard_storm, win_steps))
\end{pycon}

\smallsubsubsection{Conditional reachability probabilities~\cite{conditional}}
$\Prob_{\max=?}(F\,\varphi \mid F\,\theta)$ express the probability of reaching a $\varphi$-state, conditioned under the event that a $\theta$-state is reached.
We compute the maximal winning probability under the condition that the raven is eventually only one step away from the orchard.
Checking the property \code{'Pmax=? [F "PlayersWon" || F "RavenOneAway"]'} yields $0.3198$ which is significantly lower than the overall winning probability of $0.63$.

\smallsubsubsection{Multi-objective queries~\cite{DBLP:phd/dnb/Quatmann23}}
$\mathit{multi}(\Phi_1, \dots, \Phi_m)$ allow to compute possible tradeoffs between various (single-objective) properties $\Phi_1, \dots, \Phi_m$.

For the Orchard game, the multi-objective query asks for the maximal expected number of rounds of the game among all policies that induce a winning probability of at least 60\%.
\begin{pycon}
>>> query = 'multi(R{"rounds"}max=? [F ("PlayersWon" | "RavenWon")], \
...                P>=0.60 [F "PlayersWon"])'
>>> print(model_check(orchard_storm, query))
21.79468486
\end{pycon}
We see that using a nearly optimal winning strategy also reduces the expected number of rounds played---which is 22.3391 when considering all policies.

\section{Step Three: Working with Policies}
\label{sec:policies}
As outlined in the introduction, the use of modern probabilistic model checking increasingly asks for diagnostic information, \eg the policies that optimize the objectives that are being checked. We consider their extraction,  the analysis of such policies with respect to alternative objectives, and  the extraction of policies that can be concisely represented.

\subsection{Computing Policies}
\label{subsec:computing_policies}
In this tutorial, we are interested in understanding aspects of an optimal policy to win the Orchard game, as computed by \storm.
We already saw in \cref{fig:stormvogel_results} how to infer optimal choices from the visualization.
However, this is only feasible for smaller state spaces and we consider a more general approach.
First, we inform \storm that we want to compute an optimal policy as a by-product of the model checking call, by adding the argument \code{extract_schedulers=True}.
Computing policies can induce overhead due to the additional bookkeeping---we also note that this bookkeeping is not implemented for all types of properties.
The output of the model checking call is a quantitative result, which now also holds a policy. 
\begin{pycon}
>>> formula = stormpy.parse_properties('Pmax=? [F "PlayersWon"]')[0]
>>> result = stormpy.model_checking(orchard_storm, formula,
...                                 extract_scheduler=True)
>>> print(result.scheduler)
Fully defined memoryless deterministic scheduler:
model state:    choice(s)
           0    0
           1    0
           ...  ...
\end{pycon}
Policies for maximal reachability probabilities as in \code{formula} are memoryless and deterministic, \ie a mapping from states to actions.
The \storm representation mildly deviates: It represents for every state $s$ (index) the local index $x$ of the choice. These local indices encode the $x$'th choice for state $s$, using some internal ordering. Note that two choices in different states can yield the same action.
The policy can be queried to understand for every state which action is selected.
We use the high-level state information present in \code{orchard_prism}, \cf~\cref{subsec:other_inputs}, and obtain a more informative representation of the policy:
\begin{pycon}
>>> for state in orchard_prism.states:
...     choice = result.scheduler.get_choice(state)
...     action_index = choice.get_deterministic_choice()
...     action = state.actions[action_index]
...     print(f"In state {state.valuations} choose action {action.labels}")
...
In state [... & apple=4 & pear=4 & cherry=4 & plum=4 & raven=5] 
    choose action {'chooseAPPLE'}
...
In state [... & apple=3 & pear=4 & cherry=4 & plum=4 & raven=5] 
    choose action {'choosePEAR'}
...
\end{pycon}
We see that initially---when all trees are full---the player  chooses \gapple{} for the dice outcome \basket{}. After picking \gapple{}, the optimal strategy chooses \pear{} instead.

Some properties, such as reward-bounded reachability probabilities, require history-dependent schedulers, which are typically represented using a finite-state machine---also in \storm.
In this tutorial, we do not explore these policies further.
While the policy above is represented on the level of the input, it necessarily is a very large object, as it depends on the number of underlying states.
This problem is amplified for policies that require memory.
In \cref{subsec:compact_policies}, we therefore circumvent the explicit representation of policies.

\subsection{Analyzing the Induced Submodel}
\label{subsec:induced_model}
After obtaining a policy, a standard question is to analyze the MDP and policy with respect to additional properties. 
For example, we may want to calculate the probability to collect all \cherries{} before the raven arrives using the current policy and contrast this probability with a policy that optimizes collecting \cherries{}.
To support analyzing policies, we can create the induced Markov chain $\M^\pol$ by applying policy $\pol$ on MDP $\M$.
\begin{pycon}
>>> # Get induced model
>>> induced = orchard_prism.apply_scheduler(result.scheduler, True)
>>> # Analysis (allCherriesPicked means cherries = 0 and raven > 0)
>>> all_cherries = 'Pmax=? [F "allCherriesPicked"]'
>>> print(f"Prob for fixed: {model_check(induced, all_cherries)}")
>>> print(f"Prob for optimal: {model_check(orchard_prism, all_cherries)}")
Prob for fixed: 0.7726100987523447
Prob for optimal : 0.9120560407842777
\end{pycon}
We see that the policy optimized for collecting \cherries{} yields a higher probability of picking all \cherries{} than the submodel induced by the overall winning strategy.

\subsection{Compact Policies}
\label{subsec:compact_policies}
The policy in \cref{subsec:computing_policies} was represented as an explicit list.
There are various ways to compress a given policy: Via decision tree learning (available via the \dtcontrol tool~\cite{DBLP:conf/tacas/AshokJKWWY21}) one can create heuristically small trees for a policy, whereas \dtmap~\cite{DBLP:conf/cav/AndriushchenkoCJM25} (via the \paynt tool~\cite{DBLP:conf/cav/AndriushchenkoC21}) allows to find a minimal decision tree for a given policy.
Both tools allow interfacing with \storm.
By default, we can find a decision tree with 15 layers, which is not particularly helpful.
The challenge is that---for this model---there is no small decision tree for the policies that \storm generates.
The root cause here combines two challenges:
\begin{compactenum}
    \item The predicates that the decision tree is allowed to use is not adequate.
    \item The policy computed by \storm{} makes (arbitrary) choices that yield policies that cannot easily be compressed.
\end{compactenum}
\begin{listing}[tp]
	\codefile{code/paynt.py}{1}{11}
	\lstvspace
	\caption{Searching for small decision trees with \paynt{}}
	\label{listing:paynt:dtpaynt}
\end{listing}
Solving either of these challenges is insufficient. In~\cref{listing:paynt:dtpaynt}, we manually add adequate predicates and run a computationally expensive decision tree synthesis task in \paynt{}~\cite{DBLP:conf/cav/AndriushchenkoCJM25} to find policies that satisfy a property and are representable as a small tree. By enforcing predicates for each fruit type that express that the type has the most remaining fruit in a tree, we find a small decision tree within a minute: One should pick \gapple{} if there are more \gapple{} than other remaining fruit, or otherwise \pear{} if they are more \pear{} than others, etc.

\section{Step Four: MDPs with More Uncertainty}
\label{sec:uncertainty}
While MDPs are the standard model to describe processes that are subject to both nondeterminism and probabilistic uncertainty, not every process can adequately be represented as an MDP.
We briefly show how \storm{} can go beyond the verification of MDPs by means of imprecise probabilities and partial observations, while noting that support for such extensions is still actively developed.

\subsection{Imprecise Probabilities}
\label{subsec:imdp}
Transition probabilities in MDPs are often an abstraction of more complicated processes, such as dice rolls in the Orchard game.
Instead of precise point estimates for probabilities, \storm{} supports interval estimates in two flavors.

\smallsubsubsection{Interval MDPs.}
\emph{Interval MDPs (iMDPs)} replace the precise probabilities in MDPs with intervals of possible probabilities~\cite{DBLP:journals/sttt/BadingsSSJ23}.
For the sake of this tutorial, one can consider interval MDPs as a set of MDPs, covering all contained point estimates that yield proper successor distributions.\footnote{Interval MDPs have two semantics: \emph{static} and \emph{dynamic uncertainty}. These semantics coincide for indefinite reachability probabilities and expected rewards~\cite{DBLP:journals/mor/Iyengar05}. The static semantics used here make it easier to explain the difference to parametric MDPs.}
We replace the dice roll probabilities from the point estimates $\nicefrac{1}{6}$ to the intervals $[\nicefrac{5}{36}, \nicefrac{7}{36}]$.
This can be done by changing all occurrences of
\code{fair_dice_prob}
in \cref{lst:stormvogel_transitions_extract} with the interval
\code{Interval(fair_dice_prob-(1/36), fair_dice_prob+(1/36))}.
No further change is needed, and \stormvogel will build an interval MDP.

Uncertainty in interval MDPs can be resolved in two ways.
First, in the \emph{cooperative} or \emph{angelic} interpretation, the uncertainty is in favor of the policy.
Therefore, if the policy maximizes the probability to win Orchard, the dice probability will also be chosen in order to maximize the probability to win.
Second, in the \emph{robust} or \emph{demonic} interpretation, the uncertainty is resolved against the policy.
If the policy maximizes the probability to win, the dice probability will be chosen in order to minimize the policy's probability to win.

\begin{listing}[tp]
	\codefile{code/interval_mdp_check.py}{218}{226}
	\lstvspace
	\caption{Checking an interval MDP.}
	\label{lst:interval_mdp_check}
\end{listing}

We show how to check an interval MDP in \cref{lst:interval_mdp_check}.
In the cooperative setting, the player can achieve a maximal winning probability of $0.7961$, while in the robust setting, the player can only achieve a maximal winning probability of $0.4315$.
It turns out that in Orchard, modifying the dice probabilities by $\nicefrac{1}{36}$ has a much larger influence on the winning probability than the player's strategy.

\begin{figure}[t]
\begin{minipage}[t]{0.49\linewidth}\vspace{0pt}
\parbox[t][3.8cm][t]{\linewidth}{%
\begin{dmath*}
\scriptstyle
    \big({ 24 \cdot p^7 } + { 84 \cdot p^6 q } - { {80 \cdot p^6} } + { 114 \cdot p^5 q^2 } - { 268 \cdot p^5 q } + { 90 \cdot p^5 } + { 75 \cdot p^4 q^3 } - { 344 \cdot p^4 q^2 } + { 283 \cdot p^4 q } - { 30 \cdot p^4 } + { 24 \cdot p^3 q^4 } - { 211 \cdot p^3 q^3 } + { 333 \cdot p^3 q^2 } - { 85 \cdot p^3 q } + {3 \cdot p^2 q^5 } - { 62 \cdot p^2 q^4 } + { 181 \cdot p^2 q^3 } - { 85 \cdot p^2 q^2 } - { 7 \cdot p q^5 } + { 45 \cdot p q^4 } - { 35 \cdot p q^3 } + { 4 \cdot q^5 } - { 5 \cdot q^4}\big) \big/ {\big(p-1\big)^{3}}
\end{dmath*}
}
	\caption{Winning probability in the pMC.}
    \label{fig:rat_func}
\end{minipage}
\begin{minipage}[t]{0.48\linewidth}\vspace{0pt}
	\centering
    \parbox[t][3.8cm][t]{\linewidth}{%
	\includegraphics[width=4.75cm]{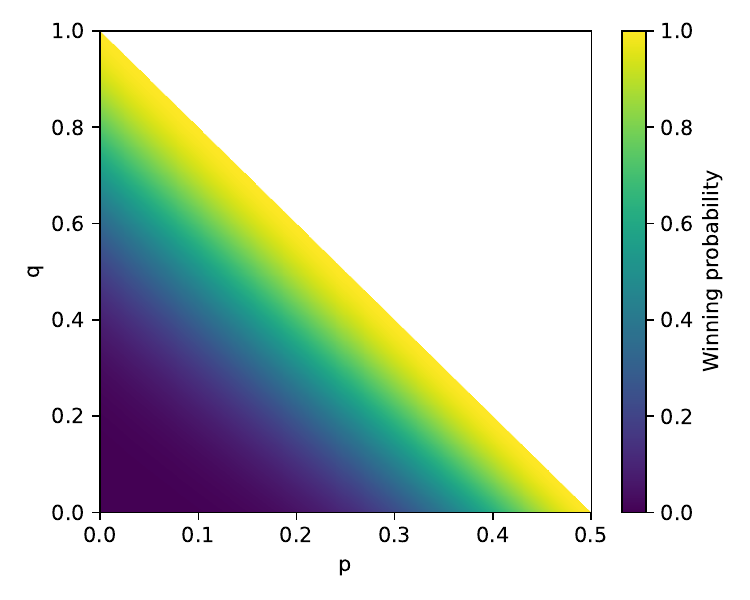}
    }
	\caption{Plot of winning probability.}
	\label{fig:plot_pmc}
\end{minipage}
\end{figure}

\smallsubsubsection{Parametric MDPs.}
The interval MDP for Orchard bounds the dice probabilities within an interval \emph{at each state}.
Different values may be optimal at different states, and thus, the dice probabilities may differ per state.
If we want to have uncertain but fixed dice probabilities, we use a \emph{parametric MDP (pMDP)}~\cite{DBLP:journals/fmsd/JungesAHJKQV24}.
We take the simplified two-fruit variant \code{orchard_simple} and parametrize it as follows: the parameter $p$ describes the probability to roll \gapple{} or \pear{}, the parameter $q$ describes the probability to roll \basket{}, and the probability to roll \raven{} is described by $1 - 2p - q$.
Similar to the iMDP, we can again replace the transition probabilities in \cref{lst:stormvogel_transitions_extract} with these functions over the parameters.
We apply the optimal winning policy from the non-parameterized game on the pMDP and obtain the induced \emph{parametric Markov chain (pMC)}.
Computing the winning probability on the pMC yields the rational function in \cref{fig:rat_func}.
The function is plotted in \cref{fig:plot_pmc}.
We can see that if \(1 - 2p - q = 0\), the player will win with probability one, and if \(p = q = 0\), the player will win with probability zero.

The rational function describing the winning probability is already very large in this simplified model with only 90 states and generally explodes.
However, \storm also exposes methods to analyze the pMC without calculating this function, such as~\cite{DBLP:journals/fmsd/JungesAHJKQV24}:
(1)~Finding feasible parameter instantiations satisfying a given property,
(2)~verifying that all instantiations in a region satisfy a given specification,
(3)~partitioning the parameter space into satisfying and violating regions,
and (4)~analyzing the pMC for monotonicity.

\subsection{Partially Observable MDPs}
\label{subsec:pomdp}
In MDP verification, we consider maximizing over all policies.
We already saw in \cref{subsec:compact_policies} that this may not always be adequate.
In particular, the policy has perfect information about the current state, which influences the choice.
In many (cyberphysical or distributed) systems, the policy should only depend on states that are known by the agent. 
Similarly, in the Orchard game, the state of the trees may not be visible to the players, however, we do assume that we know which types of fruit can still be picked.
To model this setting faithfully, the policy should not depend on the state, but instead on a set of observations---leading to a \emph{partially observable MDP (POMDP)}~\cite{DBLP:books/sp/12/Spaan12}.
\begin{definition}[POMDP]
	A \emph{POMDP} is a triple $\langle \M, Z, o\rangle$, where $M$ is an MDP with states $S$ as before, $Z$ is a finite set of \emph{observations}, and $o: S \rightarrow Z$ assigns observations to states such that $\Act(s) = \Act(s')$ implies $o(s) = o(s')$.
\end{definition}
In POMDPs, we consider observation-based, history-dependent policies which are of the form $\sigma\colon (Z \times A)^{*} \times Z \rightarrow A$\footnote{Alternative representations, \eg belief-based policies, are also used.}. Intuitively, as we must quantify over this infinite set, the verification of POMDPs is indeed undecidable~\cite{DBLP:journals/ai/MadaniHC03}; and different algorithms focus on maximizing, \eg the reachability probability or on proving bounds on these probabilities. The API therefore has no regular \texttt{model_checking} method for POMDPs.
We can print a POMDP after building it as before:
\begin{pycon}
>>> print(pomdp)
Model type:     POMDP (sparse)
States:         22469
Transitions:    44954
Choices:        29354
Observations:   546
...
Choice Labels: 	11 labels
   * nextRound -> 3749 item(s)
   * chooseCHERRY -> 2500 item(s)
   * pickAPPLE -> 3120 item(s)
   * pickCHERRY -> 3120 item(s)
   * pickPLUM -> 3120 item(s)
...
\end{pycon}
We see that the POMDP now contains information regarding the number of different observations, as well as information regarding the action names for the choices in the model (as they matter semantically). The semantics of a POMDP can be given in terms of a belief MDP, which we can partially explore and verify. We omit relevant details for the different options here, the full code is available in the supplementary material. 
\begin{pycon}
>>> belmc = BeliefExplorationModelCheckerDouble(pomdp, belexpl_options)
>>> result = belmc.check(properties[0].raw_formula, [])
>>> print(f"Result in: [{result.lower_bound}, {result.upper_bound}]")
Result in [0.63135...,0.63135...]
\end{pycon}
First, note that the result approximates the true optimum from below and above. The result here is tight, as the underlying belief MDP is sufficiently simple. 
Second, note that the result coincides with the fully observable MDP case! That is correct as a policy can track  how much fruit is available from every type. 

We can consider a modified Orchard game, where another player randomly steals fruit before the game starts. This makes the game easier, but now an observation-based policy will perform worse than the state-based policies. The verification time for the POMDP is now significantly higher than for the MDP.
\begin{pycon}
>>> mdp_res = stormpy.model_checking(pomdp_steal,winprop,
...                                        force_fully_observable=True)
>>> print(mdp_res.at(pomdp_stealing.initial_states[0]))
0.66574...
>>> belmc = BeliefExplorationModelCheckerDouble(pomdp_steal, belexpl_options)
>>> result = belmc.check(winprop.raw_formula, [])
>>> print(f"Result in: [{result.lower_bound}, {result.upper_bound}]")
Result in [0.65986...,0.65986...]
\end{pycon}

\section{Next Steps}
\label{sec:next_steps}
Above, we showcased various analysis approaches for different model types in the previous sections.
This section presents typical use cases for the Python API:
Using symbolic model representations, using preprocessing algorithms, configuring solvers, and implementing own prototypes on top of existing routines.

\smallsubsubsection{Model representation.}
So far, we have represented the MDP through an explicit (sparse) transition matrix.
Instead, we can build the MDP symbolically and represent it through \emph{decision diagrams (DD)}~\cite{DBLP:journals/sttt/KwiatkowskaNP04}.
\begin{pycon}
>>> orchard_symbolic = stormpy.build_symbolic_model(prism_program)
>>> print(orchard_symbolic)
Model type:  MDP (symbolic)
States:      22469 (88 nodes)
Transitions: 44954 (946 nodes)
...
\end{pycon}
The symbolic MDP has the same size as the sparse MDP, but instead of explicitly storing nearly 45,000 transitions, the function is represented through a DD with 946 nodes.
We can perform model checking on this symbolic representation:
\begin{pycon}
>>> formula = stormpy.parse_properties('Pmax=? [F "PlayersWon"]')[0]
>>> symbolic_result = stormpy.model_checking(orchard_symbolic, formula,
...                                          only_initial_states=True)
>>> filter = stormpy.create_filter_initial_states_symbolic(orchard_symbolic)
>>> symbolic_result.filter(filter)
>>> print("Maximal probability: {}".format(symbolic_result))
Maximal probability: 0.631356
\end{pycon}

\smallsubsubsection{Bisimulation.}
Bisimulation minimization can be applied to minimize the state space without affecting the model behavior~\cite{BK08}.
\begin{pycon}
>>> formula = stormpy.parse_properties('Pmax=? [F "PlayersWon"]')
>>> orchard_bisim = stormpy.perform_bisimulation(orchard_prism, formula,
...                     stormpy.BisimulationType.STRONG)
>>> print("Model with {} states and {} transitions".format(
...            orchard_bisim.nr_states, orchard_bisim.nr_transitions))
Model with 956 states and 2446 transitions
\end{pycon}
For the orchard model, we can reduce the state space size by $95\%$ and thereby speed-up subsequent analysis on the reduced model.

\smallsubsubsection{Model checking algorithms.}
The underlying model checking algorithms in \storm are configured through so-called \emph{environments}.
For example, we can set the precision requirement from the default value of $10^{-6}$ to $0.1$ and see how that affects the resulting probability.
\begin{pycon}
>>> env = stormpy.Environment()
>>> prec = stormpy.Rational(0.1)
>>> env.solver_environment.minmax_solver_environment.precision = prec
>>> result = stormpy.model_checking(orchard_prism, formula[0], environment=env)
>>> print(result.at(orchard_prism.initial_states[0]))
0.5815061686029693
\end{pycon}
We can also change the underlying algorithm to \eg policy iteration or optimistic value iteration~\cite{DBLP:conf/cav/HartmannsK20}.
All methods provide the same result, but their timings differ.
\begin{pycon}
>>> methods = [stormpy.MinMaxMethod.value_iteration,
...            stormpy.MinMaxMethod.policy_iteration,
...            stormpy.MinMaxMethod.optimistic_value_iteration]
>>> for m in methods:
...     env = stormpy.Environment()
...     env.solver_environment.minmax_solver_environment.method = m
...     start = time.time()
...     result = stormpy.model_checking(orchard_prism, formula[0],
...                  environment=env, extract_scheduler=True)
...     print(f"Method: {m}")
...     print(f"Result: {result.at(orchard_prism.initial_states[0])}")
...     print(f"Time:   {time.time() - start:.2}s")
Method: MinMaxMethod.value_iteration
Result: 0.6313572986959962
Time:   0.015s
Method: MinMaxMethod.policy_iteration
Result: 0.6313573066006364
Time:   0.24s
Method: MinMaxMethod.optimistic_value_iteration
Result: 0.6313576699315776
Time:   0.018s
\end{pycon}

\smallsubsubsection{Demo: Writing an LP-based MDP Model Checker.}
\label{subsec:extensions}
One further step is extending \stormpy with own algorithms.
Extending \stormpy allows to profit from existing data structure and efficient algorithms when integrating own ideas.
We exemplify this with a fundamental approach to solve MDPs using a \emph{linear program (LP)} solver, an approach that is popular due to the flexibility to add additional constraints.
The LP encoding below computes reachability probabilities.
Given a target label \(B \in \AP\),
this LP characterizes the variables \((x_s)_{s \in S}\) as follows \cite[p.\ 856]{BK08}. Minimize \(\sum_{s \in S} x_s\) such that:
\begin{compactenum}[(1)]
    \item If \(B \in L(s)\), then \(x_s =1\),
    \item If there is no path from \(s\) to any state with label \(B\), then \(x_s=0\),
    \item Otherwise, \(0 \leq x_s \leq 1\), and for all actions \(\alpha \in \Act(s)\): \[
    x_s \geq \sum_{s' \in S} \Pf(s, \alpha, s') \cdot x_{s'}.
    \]
\end{compactenum}
\begin{listing}[t]
	\codefile{code/linear_program.py}{10}{18}
	\lstvspace
	\centering
	$\vdots$
	\codefile{code/linear_program.py}{28}{41}
	\lstvspace
	\centering
	$\vdots$
	\ifthenelse{\boolean{extended}}{\caption{Encoding an LP from an MDP (extract). The full code is given in \Cref{lst:linear_program} in the appendix.}}{\caption{Encoding an LP from an MDP (extract). The full code is given in~\cite{extended}.}}
	\label{lst:linear_program_extract}
\end{listing}
\Cref{lst:linear_program_extract} shows how to create a linear program that computes the reachability probability based on this encoding.
We first compute all states that satisfy the requirements in (1) and (2) using graph algorithms provided in \stormpy.
We then construct and solve the LP by inspecting the model's transition matrix.

\section{Conclusion}
\label{sec:outlook}
This tutorial presents usage of the probabilistic model checker \storm{}, via its Python bindings.
It demonstrates how to specify an MDP, analyze it \wrt to various properties of interest and extract additional diagnostic information such as winning policies.
The tutorial focuses on MDPs and the most used properties, and purposefully omits many functionalities in \storm, in particular with respect to continuous-time models.
It focuses on using \storm{}, rather than extending it, and thus does not discuss internal data structures, nor details on effective modeling in \prism or \jani.

\smallpar{Acknowledgments.}
This tutorial builds upon tools in the \storm ecosystem, especially on \paynt and \stormvogel.
We thank Christian Hensel, Alexander Bork, and Luko van der Maas as well as all other contributors to the \storm tool.\footnote{See \href{https://github.com/moves-rwth/storm?tab=readme-ov-file\#authors}{github.com/moves-rwth/storm?tab=readme-ov-file\#authors} for a complete list.}
We thank Roman Andriushchenko, Filip Mac\'{a}k, and Milan \v{C}e\v{s}ka for the development of \paynt, especially for the API changes for this tutorial.
We thank  Pim Leerkes and Ivo Melse for their support in developing \stormvogel.

This work has been partially supported by the NWO Open Science Fund and the NWO VENI Grant UNICORN (232.211).

\smallpar{Data availability.}
The models and Jupyter notebooks of this tutorial are publicly available in the artifact at \href{https://doi.org/10.5281/zenodo.18955665}{doi.org/10.5281/zenodo.18955665}.

\bibliographystyle{splncs04}
\bibliography{references}
\ifthenelse{\boolean{extended}}{%
	\clearpage
	\appendix
\section{Appendix}
\label{appendix}
We provide the full model specification of the Orchard game in terms of the \stormvogel \bird API in \cref{app:orchard_stormvogel} and it terms of the \prism language in \cref{app:orchard_prism}.
The Python code for the LP encoding of an MDP is given in \cref{app:python_analysis}.

We refer to the artifact at \href{https://doi.org/10.5281/zenodo.18955665}{doi.org/10.5281/zenodo.18955665} which contains the models, Python code and Jupyter notebooks to follow this tutorial.

\subsection{Stormvogel Model Specification}
\label{app:orchard_stormvogel}
We provide the full model specification of the Orchard game in terms of \stormvogel in the following.
The description of the Stormvogel model is given in \cref{subsec:model_declaration}.

The data structures for the Orchard game are given in \cref{lst:stormvogel_data_structures}.
\begin{listing}[tp]
	\codefile{code/orchard_game_stormvogel.py}{5}{35}
	\lstvspace
	\caption{General data structures for the Orchard game.}
	\label{lst:stormvogel_data_structures}
\end{listing}
The full \code{Orchard} game class is given in \cref{lst:stormvogel_game_class}.
\begin{listing}[tp]
	\codefile{code/orchard_game_stormvogel.py}{38}{86}
	\lstvspace
	\caption{The Orchard game class.}
	\label{lst:stormvogel_game_class}
\end{listing}

We provide all ingredients for the MDP $\M = (S, s_0, \Act, \Pf, \AP, L)$ in the following:
\begin{compactitem}
	\item The initial state $s_0$ is defined in \cref{lst:stormvogel_initial_state}.
	\item The list of available actions $\Act(s)$ per state is given in \cref{lst:stormvogel_available_actions}.
	\item The transition function $\Pf$ is given in \cref{lst:stormvogel_transitions}.
	\item The labeling function $L$ is given in \cref{lst:stormvogel_labels}.
	\item The reward function is given in \cref{lst:stormvogel_rewards}.
\end{compactitem}
\begin{listing}[t]
	\codefile{code/orchard_game_stormvogel.py}{89}{108}
	\lstvspace
	\caption{The available actions.}
	\label{lst:stormvogel_available_actions}
\end{listing}
\begin{listing}[tp]
	\codefile{code/orchard_game_stormvogel.py}{111}{164}
	\lstvspace
	\caption{The transition function \code{delta}.}
	\label{lst:stormvogel_transitions}
\end{listing}
\begin{listing}[t]
	\codefile{code/orchard_game_stormvogel.py}{167}{174}
	\lstvspace
	\caption{The labeling $L$.}
	\label{lst:stormvogel_labels}
\end{listing}
\begin{listing}[t]
	\codefile{code/orchard_game_stormvogel.py}{177}{182}
	\lstvspace
	\caption{The reward structure.}
	\label{lst:stormvogel_rewards}
\end{listing}
\begin{listing}[t]
	\codefile{code/orchard_game_stormvogel.py}{224}{226}
	\lstvspace
	\caption{The initial state.}
	\label{lst:stormvogel_initial_state}
\end{listing}

\clearpage
\subsection{Prism Model Specification}
\label{app:orchard_prism}
We introduced a specification of the Orchard game in the Prism language in \cref{subsec:other_inputs}.
We provide the full specification of the Orchard game in the Prism language in \cref{lst:orchard_prism_1,lst:orchard_prism_2}.
The model follows the same structure as the \stormvogel model specified in \cref{app:orchard_stormvogel}.

\begin{listing}[tp]
	\codefilecustom{code/orchard-stormvogel.pm}{text}{0}{50}
	\lstvspace
	\caption{The Prism model of the Orchard game (part 1).}
	\label{lst:orchard_prism_1}
\end{listing}
\begin{listing}[tp]
	\codefilecustom{code/orchard-stormvogel.pm}{text}{52}{101}
	\lstvspace
	\caption{The Prism model of the Orchard game (part 2).}
	\label{lst:orchard_prism_2}
\end{listing}

\clearpage
\subsection{Python Code for LP Analysis}
\label{app:python_analysis}
\Cref{lst:linear_program} provides the Python code for the LP encoding using \stormpy.
The LP is described in \cref{subsec:extensions}.

\begin{listing}[tp]
	\codefile{code/linear_program.py}{10}{57}
	\lstvspace
	\caption{Encoding an LP from an MDP.}
	\label{lst:linear_program}
\end{listing}

}{%
}%

\end{document}